\documentclass[useAMS,usenatbib]{mn2e}
\usepackage{epsfig}
\newcommand{\Lsol}{L$_{\odot}$}
\newcommand{\Msol}{M$_{\odot}$}
\newcommand{\HI}{H\,{\sc {i}}~}

\newcommand{\Msold}{M$_{\odot}$\,yr$^{-1}$}
\newcommand{\Vlsr}{V$_{\rm lsr}$}
\newcommand{\Vexp}{V$_{\rm exp}$}

\newcommand{\kms}{km\,s$^{-1}$}
\newcommand{\rin}{r$_{in}$}
\newcommand{\rout}{r$_{out}$}
\newcommand{\vsh}{\hspace*{0.3cm}}
\newcommand{\sh}{\hspace*{1.cm}}

\title[The detached shell around Y CVn]{The formation of a detached shell 
around the carbon star Y CVn}
\author[Y. Libert, E. G\'erard and T. Le~Bertre]{Y. Libert$^{1}$, 
E. G\'erard$^{2}$ and T. Le~Bertre$^{1}$\\
$^{1}$LERMA, UMR 8112, Observatoire de Paris, 61 av. de l'Observatoire,
           F-75014 Paris, France\\
$^{2}$GEPI, UMR\,8111, Observatoire de Paris, 5 place J. Janssen, 
           F-92195 Meudon Cedex, France}
\begin{document}

\date{Accepted 2007 June 25. Received 2007 June 18; in original form 2007 
April 6}

\pagerange{\pageref{firstpage}--\pageref{lastpage}} \pubyear{2007}

\maketitle

\label{firstpage}

\begin{abstract}
Y CVn is a carbon star surrounded by a detached dust shell 
that has been imaged by the {\it Infrared Space Observatory} at 90 $\mu$m.
With the Nan\c{c}ay Radio Telescope we have studied the gaseous counterpart 
in the 21-cm \HI emission line. New data have been acquired and allow to 
improve the signal to noise ratio on this line. The high spectral resolution 
line profiles obtained at the position of the star 
and at several offset positions set strong constraints on the gas temperature 
and kinematics within the detached shell; the bulk of the material should 
be at $\sim$ 100--200 K and in expansion at  $\sim$ 1--2 \kms. 
In addition, the line profile at the central position 
shows a quasi-rectangular pedestal that traces an 8 \kms ~outflow of 
$\sim$ 1.0 10$^{-7}$ \Msold, stable for about 2 10$^4$ years, which
corresponds to the central outflow already studied with CO rotational lines.\\
We present a model in which the detached shell results from the slowing-down 
of the stellar wind by surrounding matter. The inner radius corresponds to the 
location where the stellar outflow is abruptly slowed down from $\sim$ 8 \kms 
~to 2 \kms ~(termination shock). The outer radius corresponds to the location 
~where external matter is compressed by the expanding shell (bow shock). 
In this model the mass loss rate of Y\,CVn has been set constant, 
at the same level of 1.0 10$^{-7}$ \Msold, ~for $\sim$ 4.5 10$^5$ years. 
The gas temperature varies from $\sim$ 1800 K at the inner limit to 
165 K at the interface between circumstellar matter and external matter.\\
Our modelling shows that the presence of a detached shell around an AGB star 
may not mean that a drastic reduction of the mass loss rate has occurred in 
the past. The inner radius of such a shell might only be the effect of 
a termination shock rather than of an interruption of the mass loss process.
\end{abstract}


\section{Introduction}

The history of the mass loss experienced by stars on the Asymptotic Giant 
Branch (AGB) is a key issue for describing the late stages of evolution 
of low and intermediate mass (1--6 \Msol) stars, as well as an important 
ingredient for the characterization of the cosmic cycle of matter. 
Determining the mass loss rates of red giants is generally based 
on modelling radio molecular line profiles (Sch{\"o}ier 2007) or 
infrared continuum energy distributions (van Loon 2007). However 
these methods are limited to the central parts of circumstellar shells, and,  
as the mass loss rates of AGB stars are variable, it has been difficult 
to establish a balance of the mass loss over the long periods of time 
that need to be considered (10$^4$--10$^6$ years). 

In principle the \HI line at 21 cm should be a useful tracer of AGB 
circumstellar environments (Le~Bertre et al. 2005). Hydrogen dominates  
their composition ($\sim$~70~\% in mass) and should be in atomic form, 
at least in the external parts of these shells (r $\geq$ 0.1 pc). 
However the 21 cm line is weak and generally contaminated by the much more 
intense galactic emission arising on the same lines of sight. 
\HI line observations of AGB stars have long been limited to Mira 
(Bowers \& Knapp 1988). However, since 2001, using the upgraded 
Nan\c cay Radiotelescope (NRT), we readdressed this issue and detected several 
objects in emission, in particular RS Cnc (G\'erard \& Le~Bertre 2003), 
EP Aqr \& Y CVn (Le~Bertre \& G\'erard 2004, Paper I), and X Her
(Gardan et al. 2006, Paper II). Recently, we also presented the results of 
a survey of 22 sources, with 18 new detections (G\'erard \& Le~Bertre 2006, 
Paper III). Using the Very Large Array (VLA), Matthews \& Reid (2007) detected 
\HI emission coincident in both position and velocity with RS Cnc, and reported
emission close to, but not unambiguously related to, EP Aqr. 
They also reported \HI emission from the circumstellar shell of R Cas. 

Generally the \HI emissions that we detect with the NRT are extended, 
indicating shell sizes $\sim$ 1 pc. In some cases, e.g. EP Aqr (Paper I), 
they reveal complex spatial and dynamic structures. The 21 cm line thus probes 
regions much larger than those that can be studied with molecular lines, which 
are limited by photo-dissociation (r $\leq$ 0.1 pc). Extended shells have 
also been observed in the far-infrared continuum emission (60 and 100 $\mu$m) 
by the Infrared Astronomical Satellite (IRAS, Young et al. 1993a). 
The spatial extension estimated from our \HI observations are comparable to, 
or possibly larger than, those obtained by Young et al. (1993a). Some of 
the infrared shells appeared detached from the central stars. The infrared 
source associated to Y CVn was reobserved by the {\it Infrared Space 
Observatory} (ISO), which clearly detected at 90 $\mu$m a detached dust shell 
surrounding the central star (Izumiura et al. 1996). These detached shells 
were interpreted as resulting from the interaction of the expelled material 
with the Interstellar Medium (ISM, Young et al. 1993b), or from a past event 
of mass loss much higher than today (Izumiura et al. 1996). However the 
far-infrared data lack crucial kinematic information.

The \HI line profiles obtained with the NRT have in general a gaussian shape,
with a Full Width at Half-Maximum (FWHM) smaller than, or equal to, the width 
of the CO rotational lines,  implying that the expansion velocity decreases 
outward (Papers I \& III). This slowing-down of the outflows indicates 
an interaction with local material, either of circumstellar or 
interstellar origin. These high-spectral-resolution \HI profiles 
provided the first direct evidence of a slowing-down of circumstellar matter 
in the external shells of red giants.

In the present paper, we revisit the case of Y CVn, because, as the ISO 
observations show images that are rather rounded, we expect that a spherical 
symmetry may apply and therefore that the interpretation will be 
easier. Also the source is relatively bright in H\,{\sc {i}}, 
with a low interstellar contamination, so that in principle \HI profiles 
of high quality are accessible. With these favorable conditions we expect 
that the \HI profile modelling will bring useful constraints on the detached 
shell characteristics, and more generally on the detached shell phenomenon. 
We have acquired new \HI data on Y CVn in order to improve those already 
presented in Paper I. 

\section{Y CVn}
\subsection{The central star}\label{star}

Y CVn is a J-type carbon star, i.e. a carbon star whose photosphere is 
enriched in $^{13}$C. The abundance ratio, 
$^{12}$C/$^{13}$C $\sim$~3.5 (Lambert et al. 1986), is close 
to the CNO-cycle equilibrium value (3-3.5, Lattanzio \& Forestini 1999).
The effective temperature is $\sim$~2760 K (Bergeat et al. 2001), so that  
atomic hydrogen should dominate over molecular hydrogen in the atmosphere 
and eventually above it (Glassgold \& Huggins 1983). However, Lambert et al. 
(1986) detected a weak 1-0 S(0) H$_2$ line at 2.22 $\mu$m, so that 
some hydrogen should still be in molecular form within the photosphere. 
The abundance analysis performed by Lambert et al. shows also 
that Y CVn may have a slightly sub-solar metallicity ($-0.2 \pm 0.2$), 
with oxygen showing the most underabundance (--0.4). 

The exact stage of evolution of Y CVn is not clear. 
J-type stars do not show an enhancement in s-process elements, suggesting
that they have not gone through thermal pulses and that 
the surface carbon enrichment is not due to a third-dredge-up event. 
For instance, they might owe their carbon-rich composition 
to an He core flash when they were still on the Red Giant Branch (Dominy 1984).

The distance derived from the {\it Hipparcos} parallax 
(4.59 $\pm$ 0.73 mas, Perryman et al. 1997) is $\sim$~218\,pc. 
This places Y CVn at 207\,pc from the Galactic Plane in a direction 
close to the Galactic North pole (b$^{\rm II}= 72^{\circ}$). We adopt 
the {\it Hipparcos} distance, but note that it might be underestimated.
Knapp et al. (2003) have re-reduced the {\it Hipparcos} data and 
give a revised parallax (3.68 $\pm$ 0.83  mas) that tends to place the source 
somewhat further from the Sun (272~pc). Also Bergeat et al. (2002) 
estimate the parallax at $\sim$ 3.85 mas (or $\sim$~260~pc). 

The K magnitude is --0.74 (Guandalini et al. 2006) which translates to 
a bolometric magnitude of 1.96 (Le Bertre et al. 2001) or $\sim$ 6200 \Lsol 
~at 218~pc. Such a luminosity places Y CVn clearly on the AGB, probably on 
the early AGB (E-AGB) rather than on the thermally pulsing AGB (TP-AGB).

The proper motion observed by {\it Hipparcos} is --2.20, in Right Ascension 
(RA), and 13.05 mas yr$^{-1}$, in Declination. 
When corrected for the solar motion towards apex, it translates to  
13.4 and  17.4 mas yr$^{-1}$ respectively, and for a distance of 218 pc 
to 14 and 18 \kms, with respect to the (solar) Local Standard of Rest (LSR). 
It corresponds to a motion, in the plane of the sky, towards North-East 
(position angle, PA = 38$^{\circ}$).

\subsection{The circumstellar envelope}\label{circenv}

Evidence for the presence of a circumstellar shell around Y CVn 
was found by Goebel et al. (1980) who detected the 11.3-$\mu$m band 
ascribed to SiC dust grains. The 1-0 rotational line of CO was first 
detected by Knapp \& Morris (1985).

Knapp et al. (1998) have obtained a high-resolution CO spectrum of the wind 
from Y CVn. They determine a LSR radial velocity, \Vlsr = 21.1 \kms, an 
expansion velocity, \Vexp = 7.8 \kms, and a mass loss rate, 
\.M = 1.1 10$^{-7}$ \Msold ~(at 218 pc). Sch\"oier et al. (2002) obtained 
similar estimates with \Vexp= 8.5 \kms ~and \.M = 1.5 10$^{-7}$ \Msold, 
as well as Teyssier et al. (2006) with  \Vexp= 6.5 \kms 
~and \.M = 1.0 10$^{-7}$ \Msold. 
The CO emission is extended with diameters, $\phi \sim 13''$ in the 1-0 
transition, and $\sim 9''$ in 2-1 (Neri et al. 1998). 
A faint asymmetry is suspected in the 2-1 data. 

An intense emission in the $^{13}$CO 1-0 line, detected by Jura et al. 
(1988), confirms the large abundance of $^{13}$C in the Y CVn outflow.
Sch{\"o}ier \& Olofsson (2000) evaluate the $^{12}$CO/$^{13}$CO abundance 
ratio to $\sim$ 2.5, a ratio larger than the equilibrium value, indicating 
a possible enrichment in $^{13}$C of the CO molecule within the outflow. 
The mass loss rates estimated from $^{12}$CO rotational line data may 
therefore be underestimated by a factor $\sim$ 1.4. We note also that the 
underabundance in oxygen, as suggested by Lambert et al. (1986), would 
directly translate into an underabundance in CO.

\begin{figure*}
\centering
\epsfig{figure=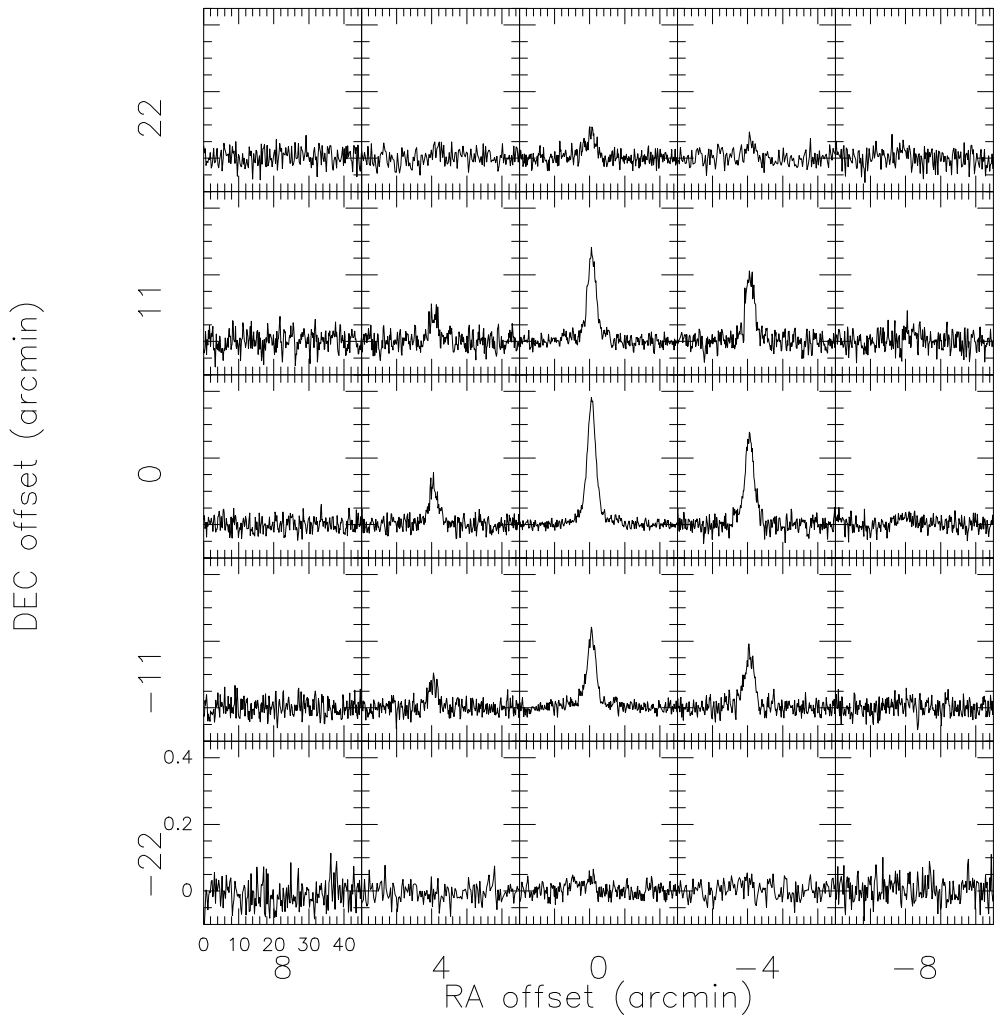,angle=270,width=20.9cm}
\caption{Map of the 21 cm \HI emission from the Y CVn circumstellar envelope 
observed with the NRT. The steps are 4$'$ in RA (1~beam) and 11$'$ 
in Declination (1/2 beam). North is up and East to the left.}
\label{Map}
\end{figure*}

An extended emission at 60 and 100 $\mu$m was discovered by IRAS (Young et al. 
1993a). This emission could be modelled with a resolved isothermal shell, 
at $\sim$ 30-40 K, of inner radius, \rin $\sim 2.8'$, and outer radius, 
\rout $\sim 5.5'$. An image at 90 $\mu$m with a better spatial resolution 
($\sim 40''$) and a better sensitivity was obtained by ISO (Izumiura et al. 
1996). It clearly shows a central infrared source surrounded by 
a quasi-circular detached shell, with \rin $\sim 2.8'$ and \rout $\sim 5.1'$. 
This image of a detached shell supports 
the modelling of the IRAS data by Young et al. (1993a). In addition, 
in the ISO image at 90 $\mu$m, the detached shell seems displaced to the West 
by about 0.5-1$'$ with respect to the central infrared source. Nevertheless, 
the circular morphology of the detached shell and the small size of the 
offset, as compared to the source diameter, leave no doubt on the association 
with Y CVn. Adopting a {\it dust-to-gas} mass ratio of 4.5 10$^{-3}$,
the mass in the detached shell can be estimated at 0.06 \Msol ~(at 218 pc).

This detached shell, observed in the dust continuum emission, is interpreted 
by Young et al. (1993b) as a product of the slowing-down of an expanding 
circumstellar shell by the surrounding ISM. On the other hand, for Izumiura 
et al. (1996), it is the result of an intense episode of mass loss, about 
two orders of magnitude larger than presently (as probed by the CO emission,  
which is limited to the central $\sim$~10$''$, Neri et al. 1998), 
with an outflow velocity $\sim$\,15\,\kms, 
that lasted about 2 10$^4$ years and stopped 1.4 10$^4$ years ago. 

In Paper I, we reported the detection of Y CVn in the \HI line at 21 cm. 
The emission is extended compared to the NRT beam size in Right Ascension 
(4$'$). The line profile is composite with a spectrally narrow component 
(Comp. 1, FWHM $\sim$ 2.9 \kms) superposed on a weak broad component (Comp. 2, 
FWHM $\sim$ 14 \kms). The 2 components are centered at \Vlsr $\sim$ 20-21 
\kms, in good agreement with the CO lines. 
Comp. 1 was found to be spatially extended ($\sim 12'$) and to be more intense 
West than East. The \HI source traced by Comp. 1 was associated to the 
detached shell observed by ISO, whereas the other one was associated with the 
Y CVn outflow already observed in CO molecular lines. 

The spatially extended emission associated to Comp. 1 and its quasi-gaussian 
line-profile of width $\sim$ 3 \kms ~were interpreted as an indication that 
the outflow velocity should decrease outward, a result which agrees with 
the Young et al. (1993b) interpretation of a slowing-down by the surrounding 
ISM. However it does not exclude a brief episode of a large mass loss 
rate in the past (see Sect.~\ref{discussion}). 

\section{Observations}\label{obs}

The NRT has a rectangular aperture of effective dimensions 
160\,m$\times$\,30\,m. At 21 cm, its beam has a 
FWHM of 4$'$ in Right Ascension (RA) and 22$'$ in Declination. The data 
on Y~CVn have been acquired in the position-switch mode with off-positions 
taken at $\pm 4'$, $\pm 8'$, $\pm 12'$, $\pm 16'$, and $\pm 32'$ 
from the central position in the East-West direction. 
The central  positions were selected on the position of the star 
(adopted from the SIMBAD database\,: 
$12^h 45^m 07.8^s +45^\circ 26' 25''$, 2000.0), and at $\pm 11'$ and 
$\pm 22'$ from it, in the North-South direction. The spectral resolution 
corresponds to 0.32 \kms. For more details we refer to Paper I.

In order to obtain the best \HI line profiles needed for our modelling, 
we have re-processed our previous data with an improved procedure (better 
weighting of the individual observations) and acquired new observations, with 
a bandwidth twice as large, allowing a better determination of the baselines.
In total, 107 hours of data have been acquired between Sept. 2002 
and Jan. 2007, and are used in the present work. 

The new map, that results from the merging of old and new data 
(Fig.~\ref{Map}), has a much better quality than the one presented 
in Paper I. The signal-to-noise ratio is on average twice better than in 
our 2004 map (Paper I). 

\begin{figure}
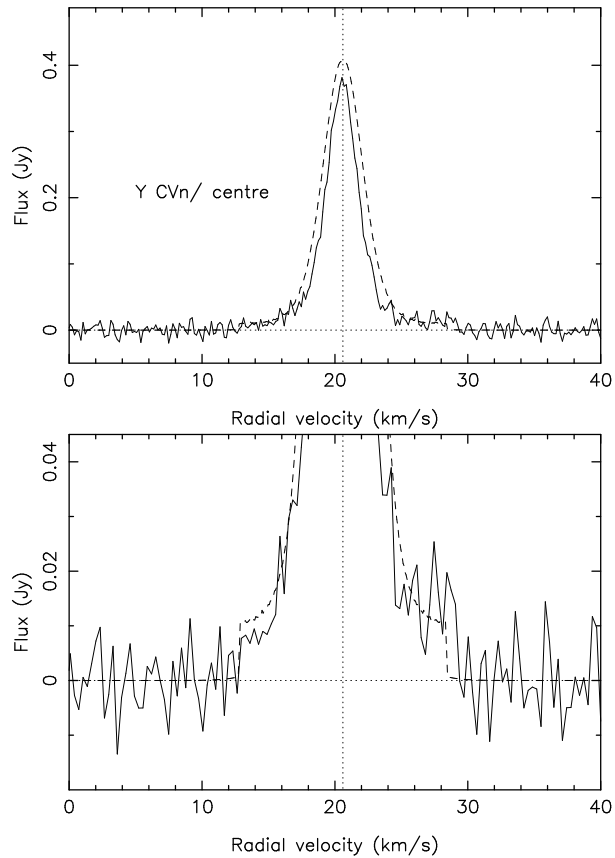

\centering
\epsfig{figure=Centre.ps,angle=-90,width=8.0cm}
\epsfig{figure=Centre_zoom.ps,angle=-90,width=8.0cm}
\caption{\HI spectrum obtained with the NRT on the stellar position 
(thin line) and modelled spectrum  (Sect.~\ref{modelling}; dashed line). 
A zoom on the lower part of the spectrum is displayed in the lower panel 
in order to show better the pedestal (Sect.~\ref{obs}).  
The vertical dotted line indicates the radial velocity 
adopted for the model (\Vlsr = 20.6 \kms).}
\label{centralpos}
\end{figure}

The \HI line at the stellar position (full line in Fig.~\ref{centralpos}) 
is clearly dominated by a narrow quasi-gaussian component (Comp.~1) of 
FWHM~$\sim$~3.1 \kms, and intensity $\sim$~360 mJy, superposed on a weak 
rectangular pedestal of width~$\sim$~16 \kms, and intensity~$\sim$~10 mJy. 
In Table~\ref{HIfit} we give the results of the line fitting by 
the sum of a gaussian and of a rectangle; as this fitting closely matches the 
data, it is not represented in the figures. The dashed lines result from the 
modelling of the whole map that will be discussed in Sect.~\ref{modelling}.
The radial velocities of the two components are nearly the same 
and in agreement, within 0.5 \kms, with that of Y~CVn as 
determined from CO data (Knapp et al. 1998). The improved quality of 
our \HI spectrum allows to see that Comp.~2 might have a rectangular profile 
rather than a gaussian one (cf. Paper I); nevertheless, for continuity, 
we keep the same notation, Comp. 1 \& 2, as in Paper I. 

\begin{figure}
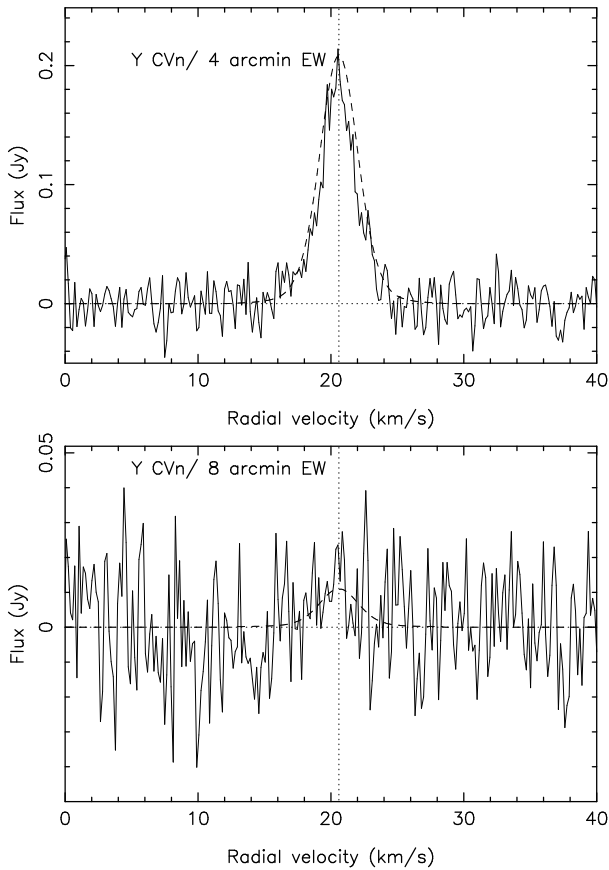

\centering
\epsfig{figure=C_1EW.ps,angle=-90,width=8.0cm}
\epsfig{figure=C_2EW.ps,angle=-90,width=8.0cm}
\caption{Upper panel: average \HI spectrum obtained with the NRT on the 
two off-positions at $\pm 4'$ in RA (thin line) and modelled spectrum  
(Sect.~\ref{modelling}; dashed line). Lower panel: same, with the 
two off-positions at $\pm 8'$.} 
\label{pos12EW}
\end{figure}

East and West of the central position (Fig.~\ref{pos12EW}, upper panel) 
the \HI line profile is essentially gaussian-like, 
i.e. with the present signal-to-noise ratio 
we do not detect a pedestal. We confirm the asymmetry between East and 
West, which is consistent with that seen on the ISO image at 90 $\mu$m. 
From the relative fluxes of Comp.~1 between the central position and the 
two offset positions, we estimate the displacement at~$\sim 1'$ West. 
As noted in Sect.~\ref{star}, this displacement is small compared to the 
source size. In addition, we note that the radial velocity of Comp.~1 is 
consistent with that of Y~CVn further supporting their association.
We do not detect the source at $\pm 8'$ from the stellar position 
(Fig.~\ref{pos12EW}, lower panel). The diameter of the \HI source can then 
be estimated at $8' \pm 4'$.

\begin{table}
\centering
\caption{Fits to the two \HI components of Y CVn (not represented in the 
figures). Comp.~1 is fitted with a gaussian profile, and Comp. 2 with a 
rectangular profile. In the case of Comp.~1 we quote the FWHM and for Comp~2 
the full width. For comparison the CO (3-2) profile obtained by Knapp et al. 
(1998) gives \Vlsr = 21.1 \kms ~and \Vexp = 7.8 \kms.}
\begin{tabular}{lccc}
\hline
                            & V$_{\rm cent}$ & Width    & F$_{\rm peak}$ \\
                            & \vsh\kms       & \vsh\kms & mJy   \\
\hline
Comp. 1 (center)            &  20.5  & 3.1  & 358. \\
Comp. 2 (center)            &  21.1  & 15.6 &  10.\\ 
Comp. 1 ($\pm 4'$ in RA)    &  20.5  & 3.2  &  179. \\ 
Comp. 1 ($\pm 11'$ in Dec.) &  20.6  & 3.1  &  232. \\ 
Comp. 1 ($\pm 11'$ in Dec.  &        &      &  \\ 
\sh     and $\pm 4'$ in RA) &  20.5  &  3.2 &  132\\ 
\hline
\end{tabular}
\label{HIfit}
\end{table}

\begin{figure}
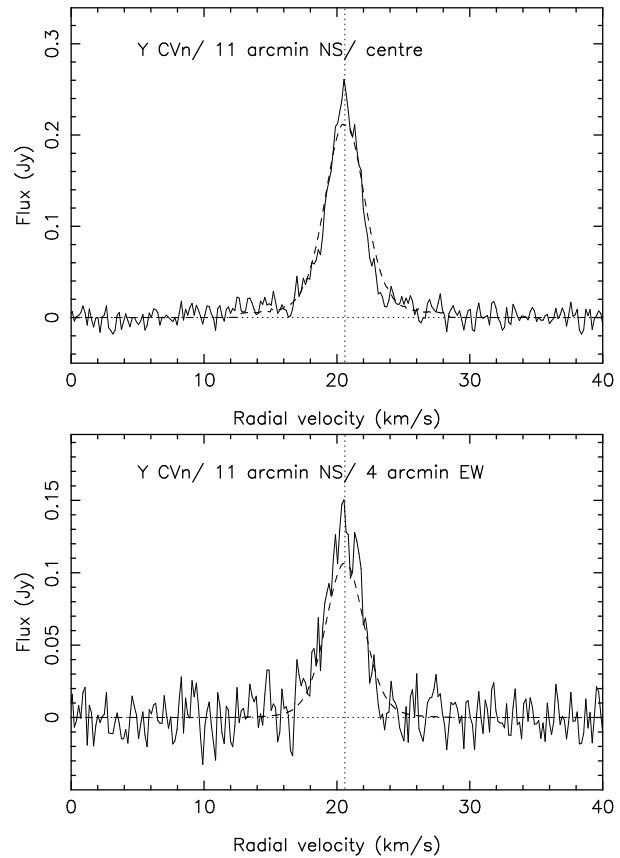

\centering
\epsfig{figure=1_2NS_Centre.ps,angle=-90,width=8.0cm}
\epsfig{figure=1_2NS_1EW.ps,angle=-90,width=8.0cm}
\caption{Upper panel: average \HI spectrum obtained with the NRT on the 
two positions at $\pm 11'$ in Declination (thin line) and modelled spectrum  
(Sect.~\ref{modelling}; dashed line). Lower panel: same, with the 
four off-positions at $\pm 11'$ in Declination and $\pm 4'$ in RA.} 
\label{posNorthSouth}
\end{figure}

The source is clearly detected at $\pm 11'$ in the North-South direction 
(Fig.~\ref{posNorthSouth}, upper panel) with more than half the flux 
on Y~CVn. Therefore it is also extended in Declination, but our spatial 
resolution is not sufficient to characterize the size in this direction. 
In the map, we note that the source is slightly more intense North than South, 
by about 10 \% at $\pm 11'$, which is an indication of a slight (relative to 
the beam) displacement North that we estimate at $1' \pm 0.5'$.
Finally, the source is also detected at the four positions $\pm 4'$ in RA 
and $\pm 11'$ in Declination (Fig.~\ref{posNorthSouth}, lower panel).

By integrating the line profiles over the full map, the total \HI flux 
associated to Y~CVn can be estimated at 3.38 Jy\,\kms. At 218 pc, 
this translates to 3.8 $\pm$ 0.2 10$^{-2}$ \Msol ~in atomic hydrogen. 
Most of it is associated to Comp.~1. We estimate the flux corresponding 
to Comp.~2, from the width and height of the pedestal detected on 
the Y~CVn position, at 0.16 Jy\,\kms, or 1.8 10$^{-3}$ \Msol.

\section{Interpretation}\label{interpretation}

Although we are limited by the spatial resolution of the NRT 
($4' \times 22'$), by combining our high-spectral resolution \HI data with 
the ISO image obtained in the dust continuum by Izumiura et al. (1996), 
we can obtain a description of the Y CVn circumstellar environment that 
takes into account the kinematics of the gas. 

The source traced by Comp. 2 can be associated with 
the central infrared source. 
Its rectangular profile is typical of an optically thin \HI emission of 
a uniformly expanding shell (Paper I). The expansion velocity ($\sim$ 8 \kms) 
is consistent with the outflow velocity obtained from CO rotational lines. 
As Comp.~2 is also centered on the radial velocity of Y CVn, we can infer 
that the corresponding source has a nearly spherical geometry.

The source traced by Comp. 1 is spatially resolved by the NRT and has a 
diameter, $\phi \sim 8' \pm 4'$. Its hydrogen mass is consistent with the 
total mass of the detached shell estimated by Izumiura et al. (1996).
Therefore it is reasonable to associate it to the infrared detached shell.  
The spectral width, $\sim$ 3 \kms, shows that 
the material is moving outwards at a reduced velocity as compared to 
the source traced by Comp. 2. This is in agreement with the analysis of 
Young et al. (1993b). 

We now pursue this analysis and consider the interaction of the Y CVn outflow 
with external matter. The infrared image and our \HI map show an offset 
of 1$'$ West with respect to the central source, but this offset is small 
compared to the diameter of the detached shell, and in the ISO image 
the detached shell has a quasi-circular appearance. Also the central velocity 
in \HI (20.6 \kms) is close to that in CO (21.1 \kms), so that no large 
deviation along the line of sight can be detected. 
In the following we assume spherical symmetry, which in particular implies 
that the external medium is at rest with respect to Y CVn.

\begin{figure}
\centering
\epsfig{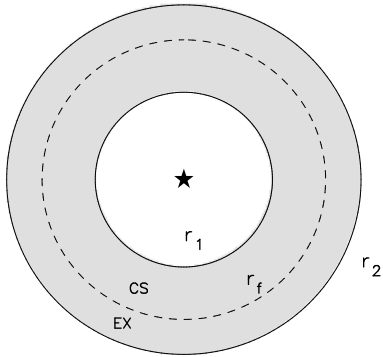}
\caption{Schematic view of the Y CVn detached shell. The termination shock 
is located in $r_1$, the contact discontinuity in $r_f$, and the bow shock 
in $r_2$. CS stands for circumstellar material and EX, for external material 
(see Sect.~\ref{interpretation}).}
\label{schema}
\end{figure}

The interaction of a spherical outflow with external matter leads to the 
formation of a region of compressed material within two spherical boundaries 
(Fig.~\ref{schema}; Lamers \& Cassinelli 1999). The internal boundary, $r_1$, 
defines the surface at which the supersonic outflow is abruptly slowed-down 
by compressed circumstellar material (termination shock). The external 
boundary, $r_2$, defines the surface at which the external medium is 
compressed by the expanding shell (bow shock). Within these two limits 
we find compressed materials of stellar (CS) and external (EX) origins 
that are separated by a contact surface defined by $r_f$. In this description 
no material is allowed to flow over the contact discontinuity in $r_f$.
Finally, inside $r_1$, the stellar outflow is in free expansion. 

We interpret the detached shell observed by ISO and the \HI source traced 
by Comp. 1 as this region of compressed material delineated by $r_1$ and 
$r_2$, and identify $r_1$ with \rin ~and $r_2$ with \rout. 
I.e. we assume that the dust and the gas have a similar spatial distribution; 
this assumption is further discussed in Sect.~\ref{discussion}. 
We also adopt the best estimates from Izumiura et al. (1996), 
so that for a distance of 218 pc, $r_1$ = 0.18 pc, and $r_2$ = 0.32 pc.

The width of the \HI emission allows to estimate an upper limit to the average 
\HI temperature within the compressed shell. For a maxwellian distribution 
of hydrogen atoms at temperature T, the \HI emission projected on a line 
of sight has a gaussian profile of width (G\'erard 1990): 
\begin{equation}       
FWHM(km~s^{-1}) = 0.214 (T(K))^{1/2}
\label{thermbroad}
\end{equation}
Therefore, the temperature of the gas within the detached shell is, 
on average, at most 210 K.
Conversely, the bulk of the material in the detached shell is moving outwards 
at a maximum velocity $\sim$\,1/2 FWHM, i.e. $\sim$\,2 \kms.

The source associated to Comp. 2 can be identified with the freely flowing 
stellar wind that fills the region inside the termination shock ($r < r_1$). 
The crossing time is about 22 10$^3$ years, which translates to an average 
mass loss rate in atomic hydrogen of about 0.8 10$^{-7}$ \Msold.

The atomic hydrogen mass of the detached shell, 
M$_{DT}$ = 3.6 10$^{-2}$ \Msol, can 
be separated in two components\,: a circumstellar part, M$_{DT,CS}$, located 
between $r_1$ and $r_f$, and an external one, M$_{DT,EX}$, between $r_f$ 
and $r_2$ (see Fig.~\ref{schema}). 
To estimate the latter we assume that it is equal to 
the hydrogen ISM mass enclosed in a sphere of radius $r_2$. 
Following Young et al. (1993b) we adopt an ISM density of 0.25 H cm$^{-3}$ 
from a galactic scale height of 100 pc, and an ISM density of 2 H cm$^{-3}$ 
in the galactic plane. In these conditions, M$_{DT,EX} \sim$ 0.9 10$^{-3}$  
\Msol, and therefore M$_{DT,CS}$ can be estimated at $\sim$ 3.5 10$^{-2}$ 
\Msol ~in atomic hydrogen. The mass 
of the detached shell appears clearly dominated by circumstellar material.
We note that Y CVn is relatively close to the Sun, in a direction for which 
the interstellar medium might be deficient (Sfeir et al. 1999), so that 
we might have been overestimating the ISM density. 

Finally, if we assume that the mass loss rate has been constant with time,  
and adopt the total mass derived from Comp. 2, M$_{DT,CS}$, 
we obtain a characteristic time 
for the formation of the detached shell of $t_{DS} \sim$ 4.5 10$^5$ years, 
$\sim$ 20 times larger than the crossing time of the freely flowing 
wind region. A characteristic velocity for the expansion of the detached 
shell is then\,: $r_1/t_{DS} \sim$ 0.4 \kms. 
This is probably an overestimate of the present expansion rate, as the 
expansion rate should decrease with time (Young et al. 1993b).

In conclusion the detached shell appears as a region of the Y~CVn 
environment where circumstellar matter has been slowed down and is stored 
for a few 10$^5$ years before being injected in the ISM.

\section{Modelling}\label{modelling}

For the freely expanding wind zone ($r < r_1$), we consider an outflow 
with a constant velocity ($v_0$~=~8~\kms) and a volumic density varying 
in $1/r^2$, corresponding to a constant hydrogen mass loss rate. 
The temperature of the gas is assumed to be low enough that its effect on 
the \HI profile can be neglected. Indeed, at distances larger than 
10$^{16}$ cm the gas temperature should be lower than 100 K (e.g. Kahane \& 
Jura 1994) and the thermal broadening should stay small compared to $v_0$.

In order to describe the detached shell around Y~CVn ($r_1 < r < r_2$) 
in simple 
terms, we make the hypotheses of stationarity and spherical symmetry; 
all quantities depend only on $r$, the distance to the central star, and 
in particular the limits $r_1$, $r_f$ and $r_2$ are treated as independent of 
time. The mass flow is constant, and the same rate apply in the detached shell 
from $r_1$ to $r_f$ as in the freely expanding wind zone ($r < r_1$):
\begin{equation}
\dot{M}=4\pi r^2 \rho v 
\label{consmass}
\end{equation}
where $\rho$ and $v$ are the density and velocity in $r$. 
For $\dot{M}$, we adopt the value corresponding to that determined 
in Sect.~\ref{interpretation}, 
$\dot{M}_{\rm H \sc I}$~=~0.8 10$^{-7}$ \Msold.

We assume that the gas behaves as an ideal gas of mean molecular weight, 
$\mu$. The pressure ($p$), density and temperature ($T$) are then related 
by the equation of state:
\begin{equation}
p = \rho \frac{kT}{\mu m_H} = \rho c^2
\label{idealgas}
\end{equation}
where $c$ is the isothermal sound velocity, $k$ the Boltzmann constant, 
and $m_H$ the mass of the hydrogen atom. For a neutral atomic gas
with 10 per cent $^4$He, and 90 per cent H, $\mu$~=~1.3.

The equation of motion for an ideal gas in spherical geometry is 
given by :
\begin{equation}
v \frac{dv}{dr} = - \frac{1}{\rho} \frac{dp}{dr} = - \frac{1}{\rho} \frac{k}{\mu m_H} \frac{d\rho T}{dr}
\label{equmotion}
\end{equation}
(i.e. we neglect the stellar gravity). Using Equ.~\ref{consmass}, which 
is valid from  $r_1$ to  $r_f$, we re-write 
(\ref{equmotion}) as:
\begin{equation}
\frac{dv}{dr}(v- \frac{c^2}{v}) = 2 \frac{c^2}{r} - \frac{k}{\mu m_H} \frac{dT}{dr}
\label{equmotion2}
\end{equation}

The properties of the gas in $r_1$ are given by the jump conditions 
(Dyson \& Williams 1997):
\begin{equation}
\rho_0 v_0 = \rho_1 v_1 
\end{equation}
\begin{equation}
\frac{v_1}{v_0} = \frac{M_0^2 + 3}{4M_0^2} \approx \frac{1}{4}
\end{equation}
where $\rho_0$, $v_0$ and $M_0$ are the upstream density, velocity 
and Mach number. 
Equ.~6 is a general expression while Equ.~7 only applies to a 
mono-atomic adiabatic gas.
From the previous discussion we adopt $v_0 \sim$ 8 \kms 
~and deduce $\rho_0$ from the mass loss rate. We estimate $M_0 \sim 20$ 
assuming an upstream temperature, $T_0$~=~20~K. 
The downstream temperature within $r_1$, $T_1$, is given by:
\begin{equation}
T_1 \approx \frac{3 \mu m_H}{16 k} v_0^2 \sim 1800 K
\end{equation}
As the average temperature within the detached shell is at most 210 K 
(previous section), the gas must cool down to a temperature lower than 
210 K within the time lapse of the detached shell formation 
($\sim$ 4 10$^5$ years). This cooling should occur through atomic lines 
and/or dust emission, and probably depends on metallicity. 
(Some heating of the gas by dust may also be expected close to $r_1$,
see Sect.~\ref{discussion}.)
We do not estimate the cooling rate, but rather adopt a temperature profile 
through the detached shell that we will constrain from our \HI observations.
In practice we adopt a law of the form: 
\begin{equation}
log~\frac{T}{T_1} = a~log~\frac{r}{r_1} 
\label{tempprofile}
\end{equation}
with $T_1$ = 1800 K. Equ. (\ref{equmotion2}) can then be re-written as:
\begin{equation} 
\frac{dv}{dr}(v- \frac{c^2}{v}) = \frac{c^2}{r} (2 - a)
\label{equmotion3}
\end{equation}

The equation of motion (\ref{equmotion3}) is solved layer by layer, 
using the classical Runge-Kutta method, 
outwards from $r_1$, under the initial conditions of density, temperature and 
velocity, $\rho_1$, $T_1$ and $v_1$. The limit $r_f$ is set 
such that the hydrogen mass between $r_1$ and $r_f$ is equal to M$_{DT,CS}$.

From $r_f$ to $r_2$, we cannot use Equ.~\ref{equmotion2}. 
As the amount of interstellar matter between r$_f$ and r$_2$ is small compared 
to the mass in the detached shell, we simply assume a $1/r^2$ dependence 
for the density in this region, with the condition that its hydrogen mass 
is equal to M$_{DT,EX}$, i.e. 0.9 10$^{-3}$ \Msol. 
The temperature is assumed constant from r$_f$ to r$_2$ (although strictly 
speaking it is expected to increase from r$_f$ to r$_2$). The velocity 
is then obtained from Equ.~\ref{equmotion}.

The velocity, hydrogen-density and temperature profiles are then used in 
an \HI emission model that we have already developed (Papers I and II). 
For each point within the detached shell, the thermal broadening is derived 
through Equ.~\ref{thermbroad}. 
In the model the \HI emission from an envelope of matter flowing radially 
from the central 
star is convolved with the response given by a telescope of rectangular 
aperture with effective dimensions 160\,m$\times$\,30\,m.
We thus calculate the \HI emission as it should be observed by the NRT 
at the central position and at different positions offset in RA  
and Declination (dashed lines in Fig.~\ref{centralpos}, \ref{pos12EW} 
and \ref{posNorthSouth}). It should be noted that in the present modelling 
the temperature profile plays a double r\^ole, first in the kinematics 
through the isothermal sound velocity (Equ.~\ref{idealgas}), 
second in the \HI profile through thermal Doppler broadening. 

\begin{figure}
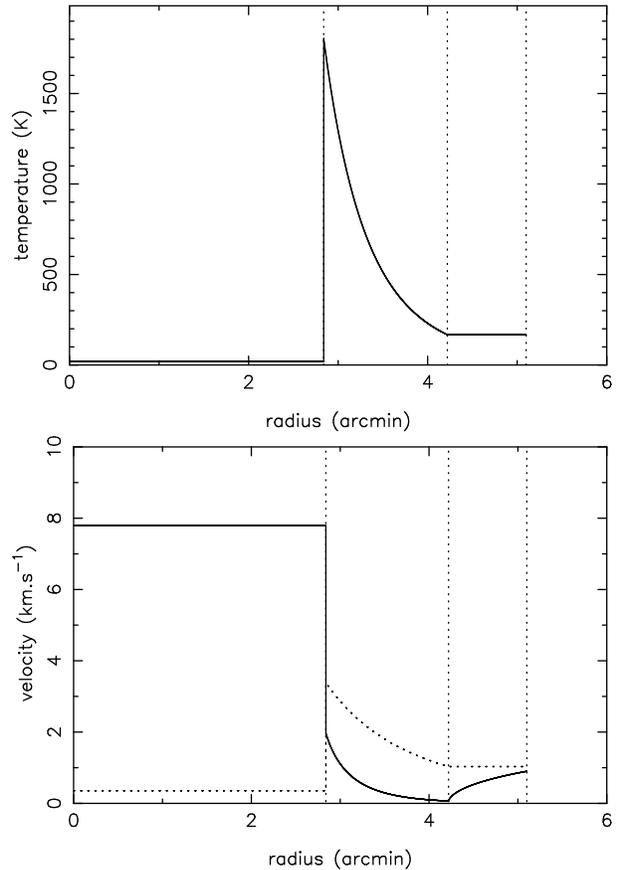
 
\epsfig{figure=proft.ps,angle=-90,width=8.0cm}
\epsfig{figure=profv.ps,angle=-90,width=8.0cm}
\caption[]{Upper panel: adopted temperature profile in the detached shell for 
the model described in Sect.~\ref{modelling}. Lower panel: velocity profile. 
The dashed line represents the isothermal sound velocity ($c$).}
\label{tempvelprofile}
\end{figure}

\begin{figure}
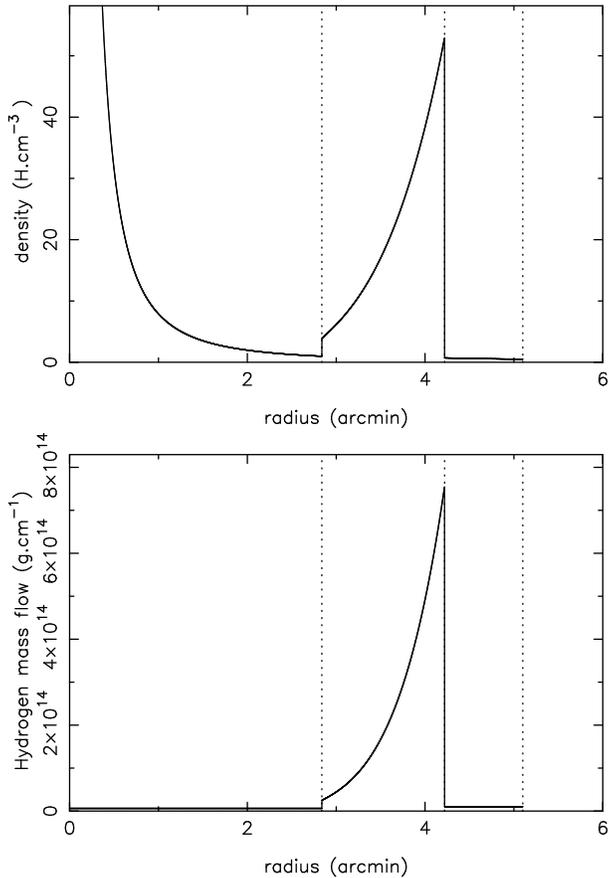
 
\epsfig{figure=profd.ps,angle=-90,width=8.0cm}
\epsfig{figure=profM.ps,angle=-90,width=8.0cm}
\caption[]{Upper panel: atomic hydrogen density profile for the model 
described in Sect.~\ref{modelling}. Lower panel: atomic hydrogen mass-flow 
profile. The vertical dotted lines mark the radii, $r_1$, $r_f$ and $r_2$, of 
the model.}
\label{densprofile}
\end{figure}

Our modelling depends only on the value selected for the constant 
$a$ in Equ.~\ref{tempprofile}. For the comparison with the observations 
(Fig.~\ref{centralpos} and following) 
we adopted a stellar velocity, \Vlsr = 20.6 \kms.
We tested several values and obtained a satisfactory fit to the observed 
\HI line profiles for $a = -6.0$. The resulting temperature profile in 
the detached shell is plotted in Fig.~\ref{tempvelprofile} (upper panel); 
it reaches a minimum in $r_f$ at $\sim$~165~K.
The corresponding velocity profile is shown in the same figure (lower panel).
Within the detached shell, the flow remains sub-sonic; the velocity profile 
that we obtain is similar to the ``breeze'' solution of the classical theory 
of stellar winds (e.g. Lamers \& Cassinelli 1999). 
The density profile and the flux of matter are shown in Fig.~\ref{densprofile}.
Both are clearly peaked at $r_f$ = 0.27 pc (4.22$'$). 

The numerical values used for the parameters of the model 
are summarized in Table~\ref{modelfit}.

\begin{figure} 
\epsfig{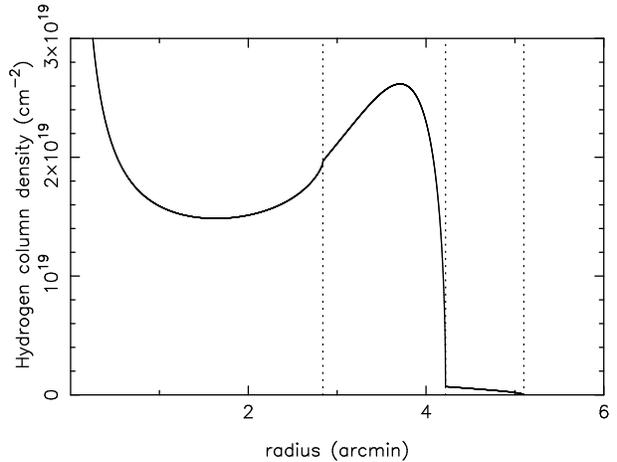}
\caption[]{Atomic hydrogen column density in the detached shell model.}
\label{columndensity}
\end{figure}

The \HI emission model developed in Papers I and II 
assumes that the brightness temperature is proportional 
to the atomic hydrogen column density (i.e. $T \ge$~10~K) and that the 
emission remains optically thin ($\tau \ll$ 1). The first hypothesis is 
clearly verified (see Fig.~\ref{tempvelprofile}, upper panel). 
The optical depth of the \HI line is given by:
\begin{equation}
\tau=\frac{3c_{light}^2}{32\pi}\frac{1}{\nu^2}A_{10}N_H\frac{h\nu}{kT}\frac{1}{\Delta\nu}
\label{opticaldepth}
\end{equation}
where $c_{light}$ is the velocity of light, $A_{10}$ the spontaneous emission 
coefficient (2.8688\,10$^{-15}$~s$^{-1}$), $N_H$ the atomic hydrogen column 
density, $h$ the Planck constant and $\Delta\nu$ the line width. Expressing 
the line width in \kms ~($\Delta V$) and the column density in cm$^{-2}$:
\begin{equation}
\tau=5.50\,10^{-19}\frac{N_H}{T\Delta V}
\label{opticaldepthAN}
\end{equation}
From Fig.~\ref{columndensity}, the column density is maximum for 
$r$ = 3.72$'$ (0.24 pc). Since $T >$~160~K and $\Delta V \sim$~3~\kms, 
the optical depth is always smaller than 0.03 across the detached shell.

\begin{table}
\centering
\caption{Model parameters (d = 218 pc).}
\begin{tabular}{ll}
\hline
\.M (in hydrogen)                 & 0.78 10$^{-7}$ \Msold\\
$\mu$                             & 1.3\\
t$_1$                             & 22\,500 years\\
t$_{DS}$                          & 427\,500 years\\
r$_1$                             & 0.18 pc (2.84$'$)\\
r$_f$                             & 0.27 pc (4.22$'$)\\
r$_2$                             & 0.32 pc (5.10$'$)\\
T$_0$($\equiv$ T$_1^-$), T$_1^+$  & 20 K, 1804 K\\
T$_f$ (= T$_2$)                   & 167 K\\
v$_0$($\equiv$ v$_1^-$), v$_1^+$  & 7.8 \kms, 1.96 \kms\\
v$_f$                             & 0.065 \kms\\
v$_2$                             & 0.9 \kms\\
n$_1^-$, n$_1^+$                  & 1.0 H\,cm$^{-3}$, 3.9 H\,cm$^{-3}$\\
n$_f^-$, n$_f^+$                  & 53. H\,cm$^{-3}$, 0.7 H\,cm$^{-3}$\\
n$_2$                             & 0.5 H\,cm$^{-3}$\\
M$_{r < r_1}$ (in hydrogen)       & 1.75 10$^{-3}$ \Msol\\
M$_{DT,CS}$   (in hydrogen)       & 3.32 10$^{-2}$ \Msol\\
M$_{DT,EX}$   (in hydrogen)       & 0.87 10$^{-3}$ \Msol\\
\hline
\end{tabular}
\label{modelfit}
\end{table}

\section{Discussion}\label{discussion}

Our modelling, that assumes stationarity and sphericity, is simplified. 
However, it allows to grasp the physical conditions within the Y CVn 
detached shell and already provides a fair adjustement to the high-resolution 
\HI spectra obtained at different positions with respect to the central star.
The fits to some line profiles appear too broad, in particular on 
the central position (Fig.~\ref{centralpos}) for which we have the highest 
signal to noise ratio. However, the total \HI flux in the map is, 
by construction, well reproduced. It could be possible to improve the fits 
by decreasing the mass loss rate in the model, but this would be at the 
expense of the agreement with the mass loss rate determined from CO 
rotational lines, and with the total \HI mass. 

In Fig.~\ref{centralpos} we note a shift in velocity of $\sim$ 1 \kms
~between the model pedestal and the observed one around 29 \kms. 
It comes from the choice in the model of a 
central velocity at 20.6 \kms, whereas the stellar velocity is more likely 
at 21.1 \kms, as can be estimated from the CO radial velocity (e.g. 
Knapp et al. 1998). As Comp. 1 is well fitted in velocity, it indicates 
that the detached shell is on average moving at a velocity of 
$\sim$ 0.5 \kms ~with respect to the central star. This could be 
an effect of a systematic motion of the star with respect to surrounding 
matter that would exert an asymmetrical pressure on the stellar outflow. 
This effect, observed along the line of sight, 
could be related to the East-West asymmetry observed in our \HI map  
(Fig.~\ref{Map}) and in the ISO image at 90 $\mu$m (Izumiura et al. 1996). 
We note that the Y CVn proper motion towards East (PA = 38$^{\circ}$, 
Sect.~\ref{star}) is consistent with this East-West asymmetry. On the other 
hand we have presently an indication for a North-South asymmetry 
that should be investigated with better spatial resolution. 

\begin{figure*}
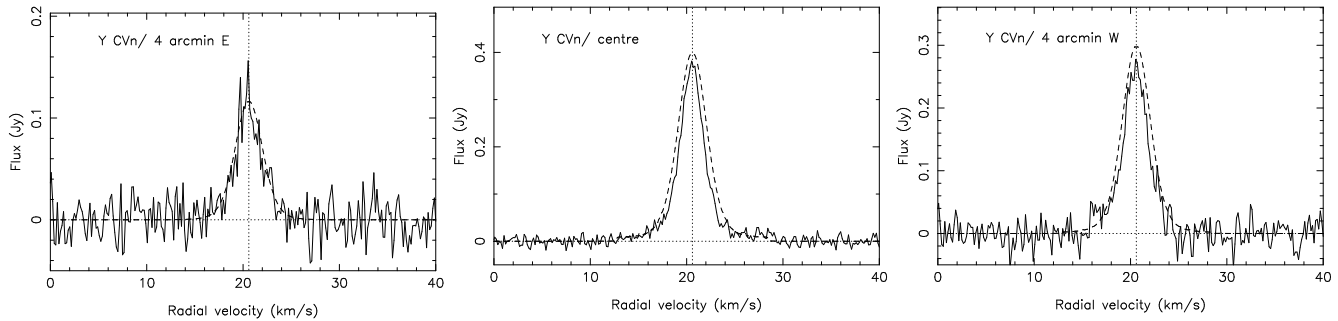
 
\centering
\epsfig{figure=C_1Est.ps,angle=270,width=5.8cm}
\epsfig{figure=Centre_non_sym.ps,angle=270,width=5.8cm}
\epsfig{figure=C_1West.ps,angle=270,width=5.8cm}
\caption[]{Effect of a shift 1$'$ West of the \HI source with respect 
to the centre of the observation grid. 
Left panel: \HI spectrum obtained 4$'$ East; 
centre panel: \HI spectrum obtained on the centre;
right panel: \HI spectrum obtained 4$'$ West. The modelled spectra 
(dashed lines) are discussed in Sect.~\ref{discussion}.}
\label{offsetmodel}
\end{figure*}

In order to test the assumption that the detached shell of Y CVn 
is displaced by 1$'$ West, 
we have performed a new modelling of the \HI profiles assuming that the 
source is 1$'$ West of the centre of our observation grid, all other 
model parameters (Table \ref{modelfit}) being kept identical. 
The results are shown in Fig.~\ref{offsetmodel}. The intensities 4$'$ East and 
West of Y CVn are well reproduced and the agreement on the central position 
is even improved. This is an indication that the displacement is real. 
In that case the hypothesis of sphericity should be re-examined and  
a non-spherical model (that might also account for the 0.5 \kms ~velocity 
shift discussed above) should be considered.

A surprising feature of our model is the relatively low density between 
$r_f$ and $r_2$ that is not confirmed by the {\it ISO} image. 
As we do not detect any \HI emission at 8$'$ East and West 
of Y CVn (Fig.~\ref{pos12EW}, lower panel), we suspect that the outer radius 
of the \HI detached shell has been overestimated by our hypothesis that 
hydrogen and dust have the same spatial distribution,
i.e. that $r_2$ = \rout = 5.1$'$. 
The displacement of the infrared source West of Y CVn may have led also 
to an overestimate of \rout ~by Izumiura et al. (1996).

In our model we adopted an empirical temperature profile. Ideally 
it would have been better to determine the temperature from a cooling 
function. This would have substantially increased the complexity of 
our modelling because the temperature enters into the equation of motion 
(Equ.~\ref{equmotion2}). Also the cooling 
function is expected to depend on metallicity which is not well constrained 
for Y CVn. In addition the mean free path of dust particles through the gas 
should be $\sim$ 10$^{9}$ cm, so that some heating of the gas due to 
friction should occur, especially close to $r_1$ where the relative velocity 
is $\sim$ 6 \kms. The grain photoelectric effect may also contribute to 
heating. The adoption of an empirical profile for the temperature includes 
all these effects. We note that a shallower profile would have led 
to a too broad \HI emission profile, and that, with a steeper one, the model 
would have failed to reproduce a correct emission on the East/West position.

The hydrogen mass loss rate that is deduced from the pedestal 
and from the inner radius of the detached shell is $\sim$ 0.8 10$^{-7}$ 
\Msold. It translates to a total mass loss rate of $\sim$ 1.0 10$^{-7}$ 
\Msold, a value compatible with those derived from CO line fitting. However, 
as noted in Sect.~\ref{circenv}, the latters are probably underestimated by 
a factor 1.4 or more. Considering the uncertainty on the CO abundance, 
a mass loss rate derived from \HI should in principle be more reliable. 
However we note that our estimate corresponds to an average over 22 10$^3$ 
years or even longer. It does not exclude possible variations on smaller 
timescales. On the other hand the CO estimates correspond to the last  
period of $\sim$ 10$^3$ years. From a detailed CO modelling 
Dinh-V-Trung \& Nguyen-Q-Rieu (2000) concluded that the Y CVn mass 
loss rate must have increased by a factor 2 or more over the last 1600 years.
Y~CVn may thus be presently in an episode of high mass loss as compared 
to the average value that we obtained in \HI. 

In our model of the detached shell we assume that the mass loss rate has been 
approximately constant for about 4-5 10$^5$ years. This may seem too large 
a duration in view of the expected time between two thermal pulses 
($\sim$~10$^5$ years or less for carbon stars, Straniero et al. 1997). However 
we noted that Y CVn is a carbon star of the rare J-type and that it may not 
necessarily be on the TP-AGB. 
On the other hand the properties that we implicitly assume for the Y CVn 
outflow fit rather well with the B-type models calculated 
by Winters et al. (2000) for carbon-rich late-type giants with stellar 
temperature $\sim$~3000~K and low variability. These models develop stationary 
outflow structures with no strong temporal variations.

Our \HI data modelling confirms that the detached shell of Y~CVn is a dense 
($\sim$ 50 cm$^{-3}$), relatively mild and quiet ($\sim$ 200 K, $\sim$ 
1--2 \kms) region where circumstellar matter stays for at least 4 10$^5$ years 
before being mixed with the ISM.

Recently, using far-infrared data from the Spitzer Space Telescope, 
Ueta et al. (2006) have detected a bow shock nebula 
with a cometary shape around the AGB star R Hya. 
Wareing et al. (2006) have modelled this nebula in terms of an interaction 
of the stellar wind with the surrounding ISM through which R Hya is moving. 
They estimate the temperature behind the shock at $\sim$ 35\,000 K, which 
corresponds to a star-ISM relative velocity of $\sim$ 35 \kms. 
At such temperature, hydrogen is expected to be entirely ionized. 

For Y CVn, although the transverse motions and the radial velocity indicate 
a motion relative to the ISM of comparable magnitude ($\sim$ 31 \kms), we find 
only a weak evidence for a distortion that could be ascribed to this motion.
We cannot exclude the presence of ionized hydrogen between the contact 
discontinuity, $r_f$, and the bow shock, $r_2$. However, we note that 
the mass in atomic hydrogen that we estimate from our \HI data, 
0.04  \Msol, agrees with the total mass estimated from the dust 
continuum  by Izumiura et al. (1996), $\sim$ 0.06 \Msol. Therefore 
the detached shell around Y CVn seems to be dominated by compressed neutral 
matter. Our modelling assumes spherical symmetry, i.e. that the surrounding 
material is at rest with respect to the central star.  
This suggests that the external matter, within which 
the detached shell is expanding, is moving together with Y CVn. 
This material might therefore not be genuine ISM, 
but possibly material remaining from an older episode of mass loss. 
We recall that the stellar evolution models (e.g. Renzini \& Fusi Pecci 1988) 
invoke an important mass loss, perhaps $\sim$ 0.2 \Msol, on the Red Giant 
Branch (RGB), that has not been detected up to now. 
In such a case our way of estimating the mass in the swept-up region 
(from $r_f$ to $r_2$), M$_{DT,EX} \sim$ 0.9 10$^{-3}$  
\Msol ~in atomic hydrogen (Sect.~\ref{interpretation}), which relies 
on the assumption that external matter is interstellar, should be revised. 
For instance an increase of M$_{DT,EX}$ by a factor 10 ($\sim$ 0.9 10$^{-2}$
\Msol) would imply a reduction of M$_{DT,CS}$ to $\sim$ 2.5 10$^{-2}$ \Msol 
~and a reduction of t$_{DS}$ to $\sim$ 3.3 10$^{5}$ years. This would 
somewhat change the structure of the detached shell in the model, 
but not fundamentally.

The phenomenon observed in Y CVn seems different from that leading to 
the formation of the detached CO shells observed around some carbon stars 
(Sch\"oier et al. 2005). These molecular shells have a typical radius $\leq$
0.1 pc and are expanding at a velocity $\geq$ 12 \kms ~(e.g. Olofsson et al. 
2000). They are smaller than the Y CVn detached shell (also they 
look thinner with a CO diameter-to-thickness ratio $\sim$ 5--10) and 
do not seem to be slowed down efficiently by an external medium, although 
there are signs of interaction with a surrounding medium. The sources 
with CO detached shells show complex CO line profiles with a double-peak 
(e.g. Olofsson et al. 1990) which is not seen in any of the Y CVn CO spectra 
(Knapp et al. 1998; Teyssier et al. 2006).
Olofsson et al. (1990) have suggested that these CO detached shells are 
produced by a mass loss eruption, possibly initiated by an He-shell flash. 
This mecanism is unlikely to play a role for Y CVn, because the central star 
is probably  still on the E-AGB.

The dynamical evolution of the interaction between stellar winds and the ISM 
has been explored by Villaver et al. (2002). They obtain transient shells 
($\sim$ 20\,000 years) associated with the wind variations induced by thermal 
pulses. Longer-lived shells could occur via interaction between successive 
events of enhanced mass loss, or via the continuous accumulation of  
ejected material in the interaction region with the ISM. 
Mattson et al. (2007) have modelled the formation of detached shells 
around TP-AGB stars. They need to combine an eruption of mass loss to 
the interaction with external matter in order to explain the properties of 
CO detached shells. They obtain thin detached shells with small 
relative thickness ($\Delta R/R \leq$ 0.01). From their modelling, 
the cases for which the observed relative thickness is large 
($\Delta R/R \geq$ 0.1) may have another origin. Therefore the mecanism 
for the formation of the Y CVn detached shell might be different from that
responsible for the molecular detached shells around carbon stars. Given 
our low spatial resolution we cannot exclude short time ($\sim$ 100 years) 
fluctuations of the Y CVn mass loss rate such as those found, for instance, 
for the carbon star IRC~+10216 (Mauron \& Huggins 2000).

On the other hand, if the scenario that we propose for Y CVn 
is correct, we expect to find similar detached shells around other AGB stars, 
and in particular around mass-losing oxygen-rich ones, because there is 
no reason for it to be specific to carbon stars. 
From our survey of atomic hydrogen around evolved stars (Paper III) we find 
\HI emission from oxygen-rich as well as from carbon-rich stars. Also the \HI 
line profiles that are observed indicate a slowing-down of circumstellar 
matter at large distance from the central stars.

\section{Prospects}\label{prospects}

The spatial resolution of our \HI data is limited to $\sim 4'$ in RA. 
Therefore we have used the ISO image at 90 $\mu$m to better constrain 
the geometry of the circumstellar shell. On the other hand the high spectral 
resolution of our \HI data has allowed to study directly the kinematics 
and the physical conditions within the Y CVn detached shell. Also the large 
collecting area  of the NRT has been useful to reach a high sensitivity on 
the low surface brightness \HI emission of Y CVn. 

However our interpretation remains limited by the low spatial resolution in 
\HI. For instance, we have used the infrared image as a guide and identified 
$r_1$ (\HI) with \rin ~(dust) ~and $r_2$ with \rout. It implicitly assumes 
that the dust and the gas spatially coincide.
Also the infrared emission 
depends on the dust temperature which should vary with distance to the central 
star. An imaging in \HI would allow to constrain directly the gas density 
within the shell as a function of the distance to the central star. 
Indeed in our model we have shown that the optical depth is always smaller 
than 1 and that the temperature in the detached shell is larger than 10 K, 
so that the \HI emission at each position should be directly proportional to 
the column density. Therefore interferometric data in \HI 
with a high sensitivity and a high spectral resolution would be very valuable 
to probe the gas physical conditions and kinematics within the detached shell 
of Y CVn.

Such high angular resolution imaging in \HI 
should allow to determine clearly $r_1$ and to compare it to \rin.
It should also allow to determine the offset of the \HI detached shell 
with respect to Y CVn. The spectrally broad component (2) is expected to be 
confined to the interior of the gas detached shell, 
which is expected to be seen only in the spectral band-width corresponding to 
the narrow component (1). Finally, we predict a concentration of hydrogen 
at $r \sim$ 0.24 pc (3.7$'$, see Fig.~\ref{columndensity}).

Other atomic lines, for instance C\,{\sc {i}} at 492 GHz, might also be 
valuable tracers of the gas within the detached shell. C\,{\sc {i}} was not 
detected by Knapp et al. (2000); however, they observed on the stellar 
position with a beam of only 15$''$ and a throw of 60$''$ ($<\,r_1$). 
In the optical range, resonant lines of Na or K could be useful. Mauron \& 
Guilain (1995) did not detect any Na\,{\sc {i}}/K\,{\sc {i}} emission, 
but similarly they looked only close to the central star (5$''$). 

The conditions for the temperature and the density that we find in our 
model might be favorable to the formation of molecular species through 
a non-equilibrium chemistry as it has been 
suggested for some regions of the diffuse ISM (Falgarone et al. 2005). 
In fact as the physical conditions within the detached shell and the 
timescale can be characterized through modelling, Y CVn could thus 
be a good target for studying this type of ISM chemistry.

\section{Conclusions}

We have presented high spectral resolution \HI data obtained on Y CVn with 
the NRT. The emission is spatially resolved with a diameter $\sim 8'\pm4'$. 
The spectrum obtained on the stellar position reveals a rectangular pedestal 
centered at +21.1 \kms ~(Comp. 2) that traces an 8 \kms ~outflow of $\sim$ 
1.0 10$^{-7}$ \Msold. This outflow corresponds fairly well to the wind 
detected in the CO rotational lines.

The spectrally narrow component (Comp. 1) is centered at +20.6 \kms ~and 
traces an expanding gas shell that we associate with the detached shell imaged 
by ISO in dust continuum emission at 90 $\mu$m (Izumiura et al. 1996). 
The bulk gas temperature within this shell is $\sim$ 100--200 K and the 
expansion velocity $\sim$ 1--2 \kms. We have developed a simplified model 
in which the detached shell is produced by a slowing-down of the Y CVn 
outflow by external matter. In this model the mass loss  is taken  
constant for $\sim$ 5 10$^5$ years with the same rate as presently, i.e. 
1.0 10$^{-7}$ \Msold. The sharp decrease of the velocity by a factor $\sim 4$ 
is interpreted as due to a shock at the inner boundary (termination shock). 
The Y~CVn detached shell thus appears to act as a lock chamber where 
circumstellar matter is stored for a few 10$^5$ years before being 
injected in the ISM.

We note that in our \HI model of the detached shell, we have adopted the 
same dimensions as those of the dust detached shell. These dimensions might 
not apply exactly and a high angular resolution mapping at 21 cm is needed. 
It would also provide the exact position of the detached shell 
with respect to Y~CVn.

We find that the surrounding medium which slows down the outflow is almost 
at rest with respect to Y CVn. Although we cannot exclude that this external 
medium is local ISM, we suggest that it might rather be made of  material left 
over from an older episode of mass loss, perhaps when the star was on the RGB.

Finally, 
we caution that the mere presence of a detached shell around an AGB star is 
not a proof that the central star has undergone an enhanced episode of mass 
loss. We stress the importance of acquiring high spectral resolution data 
to complement the dust continuum images obtained in the far-infrared.

\section*{Acknowledgments}
The Nan\c{c}ay Radio Observatory is the Unit\'e scientifique de Nan\c{c}ay of 
the Observatoire de Paris, associated as Unit\'e de Service et de Recherche 
(USR) No. B704 to the French Centre National de la Recherche Scientifique 
(CNRS). The Nan\c{c}ay Observatory also gratefully acknowledges the financial 
support of the Conseil R\'egional de la R\'egion Centre in France. 
We thank the referee and L. Matthews for valuable comments that helped us 
to improve this paper.
We thank Jean Borsenberger for developing a new and efficient procedure 
for processing NRT data, and Jer\^ome Pety for his advices in 
using the GILDAS environment. This 
research has made use of the SIMBAD database, operated at CDS, Strasbourg,
France and of the NASA's Astrophysics Data System.

\label{lastpage}


\begin{thebibliography}{}
\bibitem[]{}
Bergeat J., Knapik A., Rutily B.,  2001, A\&A, 369, 178
\bibitem[]{}
Bergeat J., Knapik A., Rutily B.,  2002, A\&A, 390, 967
\bibitem[]{}
Bowers P. F., Knapp G. R., 1988, ApJ, 332, 299
\bibitem[]{}
Dinh-V-Trung, Nguyen-Q-Rieu, 2000,  A\&A, 361, 601
\bibitem[]{}
Dominy J. F., 1984, ApJS, 55, 27
\bibitem[]{}
Dyson J. E., Williams D. A., 1997, "The physics of the interstellar 
medium", 2$^{nd}$ edition, Institute of Physics Publishing
\bibitem[]{}
Falgarone E., Hily-Blant P., Pineau des For\^ets G., 2005, Proc. ``{\it 
The Dusty and Molecular Universe''}, A. Wilson (ed.), ESA SP--{\bf 577}, p.~75
\bibitem[]{}
Gardan E., G\'erard E., Le~Bertre T., 2006, MNRAS, 365, 245 (Paper II)
\bibitem[]{}
G\'erard E., 1990, A\&A, 230, 489
\bibitem[]{}
G\'erard E., Le~Bertre T., 2003, A\&A, 397, L17
\bibitem[]{}
G\'erard E., Le~Bertre T., 2006, AJ, 132, 2566 (Paper III)
\bibitem[]{}
Glassgold A. E., Huggins P. J., 1983, MNRAS, 203, 517
\bibitem[]{}
Goebel J. H., et al., 1980, ApJ, 235, 104
\bibitem[]{}
Guandalini R., Busso M., Ciprini S., Silvestro G., Persi P., 2006, 
A\&A, 445, 1069
\bibitem[]{}
Izumiura H., Hashimoto O., Kawara K., Yamamura I., Waters L. B. F. M.,
1996, A\&A, 315, L221
\bibitem[]{}
Jura M., Kahane C., Omont A., 1988, A\&A, 201, 80
\bibitem[]{}
Kahane C., Jura M., 1994, A\&A, 290, 183
\bibitem[]{}
Knapp G. R., Crosas M., Young K., Ivezi\'c \v{Z}., 2000, ApJ, 534, 324
\bibitem[]{}
Knapp G. R., Morris M., 1985, ApJ, 292, 640
\bibitem[]{}
Knapp G. R., Pourbaix D., Platais I., Jorissen A., 2003, A\&A, 403, 993
\bibitem[]{}
Knapp G. R., Young K., Lee E., Jorissen A., 1998, ApJS, 117, 209
\bibitem[]{}
Lambert D. L., Gustafsson B., Eriksson K., Hinkle K. H., 1986, ApJS, 62, 373
\bibitem[]{}
Lamers J. G. L. M., Cassinelli J. P., 1999, ``Introduction to Stellar Winds'', 
Cambridge University Press
\bibitem[]{}
Lattanzio J., Forestini M., 1999, IAUS 191, 31
\bibitem[]{}
Le~Bertre T., G\'erard E., 2004, A\&A, 419, 549 (Paper I)
\bibitem[]{}
Le~Bertre T., G\'erard E., Winters J. M., 2005, Proc. ``{\it The Dusty 
and Molecular Universe''}, A. Wilson (ed.), ESA SP--{\bf 577}, p.~217
\bibitem[]{}
Le Bertre T., Matsuura M., Winters J. M., Murakami H., Yamamura I., 
Freund M., Tanaka M., 2001, A\&A, 376, 997 
\bibitem[]{}
Matthews L. D.,  Reid M. J., 2007, AJ, 133, 2291
\bibitem[]{}
Mattsson L., H\"ofner S., Herwig F., 2007, A\&A, in press 
(arXiv: 0705.2232)
\bibitem[]{}
Mauron N., Guilain C., 1995, A\&A, 298, 869
\bibitem[]{}
Mauron N., Huggins P.J., 2000, A\&A, 359, 707
\bibitem[]{}
Neri R., Kahane C., Lucas R., Bujarrabal V., Loup C., 1998, A\&AS, 130, 1
\bibitem[]{}
Olofsson H., Bergman P., Lucas R., Eriksson K., Gustafsson B., 
Bieging J. H., 2000, A\&A, 353, 583
\bibitem[]{}
Olofsson H., Carlstr\"om U., Eriksson K., Gustafsson B., 
Willson L. A., 1990, A\&A, 230, L13
\bibitem[]{}
Perryman M. A. C., et al., 1997, A\&A, 323, L49
\bibitem[]{}
Renzini A., Fusi Pecci F., 1988, ARA\&A, 26, 199
\bibitem[]{}
Sch\"oier F. L., 2007, Proc. ``{\it Why Galaxies Care About AGB Stars''}, 
F. Kerschbaum, C. Charbonnel \& R. Wing (eds.), ASP Conf. Ser., in press 
\bibitem[]{}
Sch{\" o}ier F. L., Lindqvist M., Olofsson H., 2005, A\&A, 436, 633
\bibitem[]{}
Sch\"oier F. L., Olofsson H., 2000, A\&A, 359, 586
\bibitem[]{}
Sch\"oier F. L., Ryde N., Olofsson H., 2002, A\&A, 391, 577
\bibitem[]{}
Sfeir D. M., Lallement R., Crifo F., Welsh B. Y., 1999, A\&A, 346, 785
\bibitem[]{}
Straniero O., Chieffi A., Limongi M., Busso M., Gallino R., 
Arlandini C., 1997, ApJ, 478, 332
\bibitem[]{}
Teyssier D., Hernandez R., Bujarrabal V., Yoshida H., Phillips T. G., 
2006, A\&A, 450, 167
\bibitem[]{}
Ueta T., Speck A. K., Stencel R. E., et al., 2006, ApJ, 648, L39
\bibitem[]{}
van Loon J. Th., 2007, Proc. ``{\it Why Galaxies Care About AGB Stars''}, 
F. Kerschbaum, C. Charbonnel \& R. Wing (eds.), ASP Conf. Ser., in press
\bibitem[]{}
Villaver E., Garc\'ia-Segura G., Manchado A., 2002, ApJ, 571, 880
\bibitem[]{}
Wareing C. J., Zijlstra A. A., Speck A.K., et al., 2006, MNRAS, 372, L63
\bibitem[]{}
Winters J. M., Le Bertre T., Jeong K. S., Helling Ch., Sedlmayr E., 2000, 
A\&A, 361, 641
\bibitem[]{}
Young K., Phillips T. G., Knapp G. R., 1993a, ApJS, 86, 517
\bibitem[]{}
Young K., Phillips T. G., Knapp G. R., 1993b, ApJ, 409, 725
\end{thebibliography}
\end{document}